\documentclass{article}
\usepackage{graphicx}
\usepackage{mathtools}
\usepackage{amsmath}
\usepackage{amssymb}
\usepackage{dsfont}
\usepackage{bbold}
\usepackage[a4paper, left=2cm, right=2cm, top=2cm, bottom=2cm]{geometry}
\usepackage{caption}
\usepackage{subcaption}
\usepackage{float}
\usepackage[normalem]{ulem}
\usepackage{microtype}
\usepackage{xargs}
\usepackage{xcolor}
\usepackage[colorinlistoftodos, prependcaption]{todonotes}

\usepackage[backend=biber, sorting=none, citestyle=numeric-comp, bibstyle=ieee, url=false, doi=false, isbn=false, eprint=false]{biblatex}
\AtEveryBibitem{%
  \clearfield{number} 
  \clearfield{note} 
  \clearfield{address} 
  \clearfield{month}
  \clearfield{day}
  \clearfield{doi}
}
\title{
The different localisation properties of the eigenmodes of the Laplacian and adjacency matrix of 2D random geometric graphs
}

\author{Luca Schaefer and Barbara Drossel}
\date{Institute for Condensed Matter Physics, Technical University of Darmstadt,
 Hochschulstraße 6, 64289 Darmstadt, Germany}

\begin{document}
\maketitle
\abstract{
We compare the spectrum and the localisation properties of the eigenmodes of the Laplacian and the adjacency matrix of 2D random geometric graphs, using numerical diagonalization of these matrices for different system sizes and connectivities. For sufficiently large ensembles of systems, we evaluate the spectrum, the probability distribution of the participation ratio and the relation between participation ratios and eigenvalues. While all eigenmodes of the adjacency matrix are localised for sufficiently large system sizes, the Laplacian matrix always leads to a small proportion of system-spanning modes due to a conservation law, and therefore to power-law tails in the probability distribution of the participation ratio and its relation to the eigenvalues. By disentangling the effects of finite system size, of mean degree, of component size distribution, and of network motifs, we provide a thorough understanding of the data.

}
\section{Introduction}
Systems with many interacting agents are often modelled as networks in which the agents are nodes and two agents are connected by a link if they interact.
Prominent examples are social networks, power grids, and transport networks~\cite{newmanNetworksIntroduction2010}.
The spectra of the matrices associated with these networks, like the adjacency or Laplacian matrix, contain information on the structure of and dynamics on these networks~\cite{vanmieghemGraphSpectraComplex2023}.
General insights into such systems are obtained by studying generic models where ensembles of networks are generated by connecting nodes according to probabilistic rules such that the networks show certain features found in the real system.
The null model where all  links between pairs of nodes are equally likely, is the Erd\H{o}s-Rényi model~\cite{erdos_renyi_1959}.
For networks in which the degree distribution is close to a power law, the Barabási-Albert model, which includes preferential attachment, is suitable~\cite{barabasiEmergenceScalingRandom1999}.
For many networks, spatial distances play an important role at determining the links. 
Examples are spatial networks like interacting wireless sensors described by the Gilbert disc model~\cite{gilbertRandomPlaneNetworks1961}, networks of ecological habitats, transport networks, or power grids (see~\cite{barthelemySpatialNetworks2011a} for a review).
They are often represented as random geometric graphs (RGG)~\cite{penroseRandomGeometricGraphs2003} or in the equivalent formulation in continuum percolation theory (see for example~\cite{balbergPrinciplesTheoryContinuum2020}). 
These RGGs are generated by placing nodes in space via a Poisson point process and connecting nodes that are sufficiently close to each other.
In contrast to the spectra of other random graphs like the Erd\H{o}s-Rényi or the Barabási-Albert model, those of RGGs have a non-vanishing discrete part due to orbits, which are sets of nodes that share the same neighbourhood~\cite{nyberg2015mesoscopic,dettmannSymmetricMotifsRandom2018}.
Dettmann and Knight computed that in two dimensions there is a maximum density of orbits at a specific connectivity radius, with the density decreasing towards zero with increasing connectivity radius~\cite{dettmannSymmetricMotifsRandom2018}.
This is especially of interest in the investigation of localisation on RGGs since some eigenvectors of the adjacency and Laplacian matrix localise on these orbits~\cite{nyberg2015mesoscopic, dettmannSymmetricMotifsRandom2018, schaeferLocalizedDelocalizedModes2025}.
The localisation behaviour of eigenvectors has also extensively been studied in the context of Anderson localisation~\cite{kramerLocalizationTheoryExperiment1993}, where the disorder of the graph resides on the diagonal entries of the Hamiltonian instead of the number of entries per row as it is for the RGGs.
There are also modifications of the original Anderson model, in which the weights of the edges are random, i.e. off-diagonal disorder~\cite{eilmesExponentsLocalizationLengths2001a,krishnaUniversalDysonSingularity2021,theodorouExtendedStatesOnedemensional1976,dysonDynamicsDisorderedLinear1953,pendryOffdiagonalDisorder1D1982,soukoulisOffdiagonalDisorderOnedimensional1981}. It is known that in the Anderson model and its modification all eigenvectors are localised in 1D and 2D systems, with 2 being the critical dimension above which there is a phase transition at a critical disorder strength ~\cite{abrahamsScalingTheoryLocalization1979,kramerLocalizationTheoryExperiment1993}. The same must hold on RGGs, where the disorder resides in the local number of neighbours instead of the weight of edges or on-site terms. 
Although the Laplacian matrix and the adjacency matrix of RGGs describe the same network, their spectrum can be expected to be qualitatively different. The reason is that the Laplacian matrix is relevant for diffusive dynamics on the RGG, which has a conserved quantity. Therefore, there must be relaxing eigenmodes of the dynamics that span the entire system, even in 1D and 2D.  Recently, we have evaluated both the spectrum of the Laplacian and adjacency matrix in one-dimensional RGGs, confirming the expectation that the two spectra and the associated localisation properties are qualitatively different. Since the number of system-spanning  eigenvectors of the Laplacian matrix increases only with the square root of system size in one dimension~\cite{hirotsugamatsudaLocalizationNormalModes1970,dunlapAbsenceLocalizationCertain1989,schaefer2025scaling,schaeferLocalizedDelocalizedModes2025}, the existence of system-spanning modes does not contradict the insight that there is no phase transition to delocalisation in 1D RGGs. In two dimensions, it is much more difficult to obtain reliable insights into the localisation properties of the eigenvectors because the linear system sizes accessible to numerical simulations are much smaller than in 1D and because eigenvectors localise less well in 2D, which is the critical dimension, leading to much larger localisation lengths and stronger finite-size effects. The literature on the spectra of 2D RGGs has so far been silent on the question how data based on the Laplacian and adjacency matrix differ qualitatively and how the conserved quantity related to the Laplacian matrix affects the extent of localisation of the eigenmodes. In this paper, we will focus on the spectra and extent of localisation (as measured by the participation ratio) obtained for the Laplacian and adjacency matrix of 2D RGGs. By disentangling the effects of system size, mean degree, component size distribution, and network motifs, and combining them with the insights gained from the 1D model we will provide an extensive understanding of these systems. 

\section{Model}
The starting point of the models under consideration is a Random Geometric Graph (RGG).
We define it by placing $N$ nodes on the unit square $[0,1]\times[0,1]$.
The coordinates are drawn independently from a uniform distribution.
Two nodes are connected by a link if their Euclidean distance is smaller or equal to a connectivity radius $r$. This means that a node is connected to all nodes that lie within the disk of area $r^2\pi$ around it, resulting in the mean degree $z=N\pi r^2$.
We choose periodic boundary conditions such that the unit square is equivalent to a two-dimensional torus.
The adjacency matrix, $A$, of a graph is defined as $A_{ij}=1$ if there is a link between the nodes $i$ and $j$, and $0$ otherwise.
Therefore, it is an $N\times N$ matrix.
The Laplacian matrix, $L$, is defined as $L\coloneqq D - A$ where $D$ is the degree matrix with $D_{ij}=\delta_{ij}k_j$ and $k_j$ is the degree of node $j$ with the Kronecker delta $\delta_{ij}=1$ if $i=j$ and $0$ otherwise.
Both matrices are symmetric and hence have real eigenvalues.
For a regular square lattice ($z=4$), the spectrum of the adjacency matrix is symmetric around $0$ since the graph is bipartite (see Fig.~\ref{fig:DOS}). This is no longer correct for an RGG. 
The spectrum of the Laplacian matrix is non-negative and the multiplicity of the eigenvalue $0$ corresponds to the number of connected components in the underlying graph~\cite{moharLAPLACIANSPECTRUMGRAPHS}.
In the following, we call the model describing the RGG by means of the negative adjacency matrix the adjacency model (AM) and the one with the Laplacian matrix Laplacian model (LM).

The eigenmodes of the AM satisfy a similar equation as the Anderson model with off-diagonal disorder and zero on-site potentials (see for example~\cite{eilmesExponentsLocalizationLengths2001a,eggarterSingularBehaviorTightbinding1978,theodorouExtendedStatesOnedemensional1976})
\begin{align}
    E\psi_{n} = -\sum_{j\in \mathcal{N}_n}\psi_j
\end{align}
with the difference that now disorder resides in the number of neighbours and not in the coupling strengths. Here, $\mathcal{N}_n\coloneqq \{i \in V: \{n,i\} \in E\}$ with the set of nodes $V$ and the set of undirected links $E$ defines the neighbourhood of node $n$. 

The eigenvalue equation of the LM 
\begin{align}
    E\psi_{n} = \sum_{j\in \mathcal{N}_n}(\psi_n - \psi_j)
\end{align}
is equivalent to the stationary equation for a system of coupled harmonic oscillators with the same mass and a coupling constant of unity. It is also equivalent to the eigenvalue problem of the discrete diffusion equation (with $-E$ being equivalent to the relaxation constant of the eigenmode), implying that the dynamics of the system has a conserved quantity $\sum_n \psi_n$.

Motivated by these analogies, we adopt the terminology from the literature on Anderson localisation. We refer to the spectrum of graph matrices as the density of states of the system and to the eigenvalues as energies.

On a two-dimensional regular lattice with $z=4n_\perp(n_\perp+1)$, where $n_\perp\in\mathbb{N}$ is the number of neighbours along the positive $x$-direction, a plane wave ansatz gives the dispersion relation of the AM
\begin{align}
    E(k) = -2\sum_{n=1}^{n_\perp}[\cos(nk_x) + \cos(nk_y)] - 4\sum_{m,l=1}^{n_\perp}\cos(mk_x)\cos(lk_y)
\end{align}
and that of the LM
\begin{align}
    E(k) &= z- 4n_\perp + 4\sum_{n=1}^{n_\perp}\left[\sin^2\left(n\frac{k_x}{2}\right) + \sin^2\left(n\frac{k_y}{2}\right)\right]- 4\sum_{m,l=1}^{n_\perp}\cos(mk_x)\cos(lk_y) \\
    &\overset{|k|\ll1}{\propto}|k|^2\propto \frac{1}{\lambda^2}\, .
\end{align}
The second sum takes diagonal edges into account, e.g. if $z=8$ or $z=24$.
The dispersion relation of the LM is shifted by $z$ along the energy axis.
In the AM, the $k$ values for $E=0$ and $z=4$ lie on a square in the band centre, whereas $|k|=0$ on the band edge in the LM.
For $z\neq4$ the lattice is not bipartite and the spectrum not symmetric around $E=0$ (see Fig.~\ref{fig:DOS}(a)).

A pecularity of RGGs is the occurrence of motifs, i.e. recurring subgraphs.
A special case are so-called orbits~\cite{nyberg2015mesoscopic}.
These are sets of nodes with the same neighbourhood.
They give rise to eigenvectors which are fully localised on these orbits.
The corresponding eigenvalues are $1$ or $0$ in the adjacency spectrum depending on whether the nodes with the same neighbourhood themselves are connected or not.
In the spectrum of the Laplacian matrix, this becomes $k+1$ and $k$, respectively, where $k$ is the degree of the nodes considered~\cite{schaeferLocalizedDelocalizedModes2025, nyberg2015mesoscopic}.

\begin{figure}[H]
    \centering
    \begin{subfigure}[b]{0.49\linewidth}
        \includegraphics[width=\textwidth]{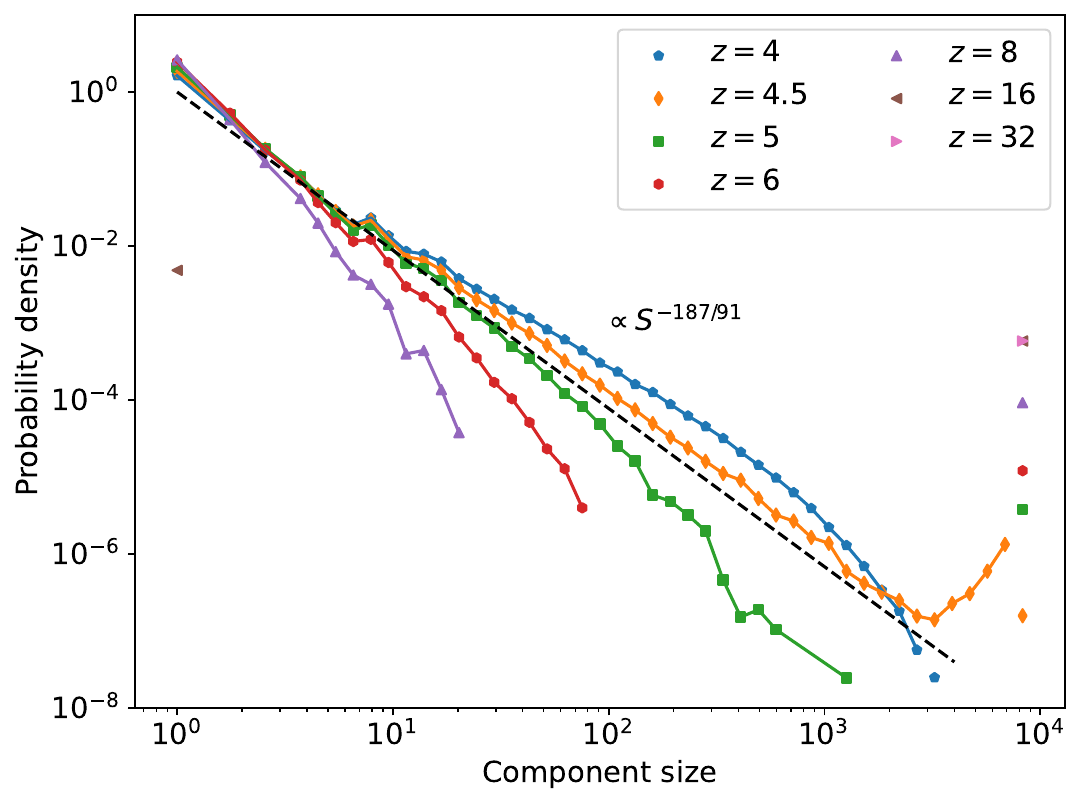}
    \end{subfigure}
\hfill
    \begin{subfigure}[b]{0.49\linewidth}
        \includegraphics[width=\textwidth]{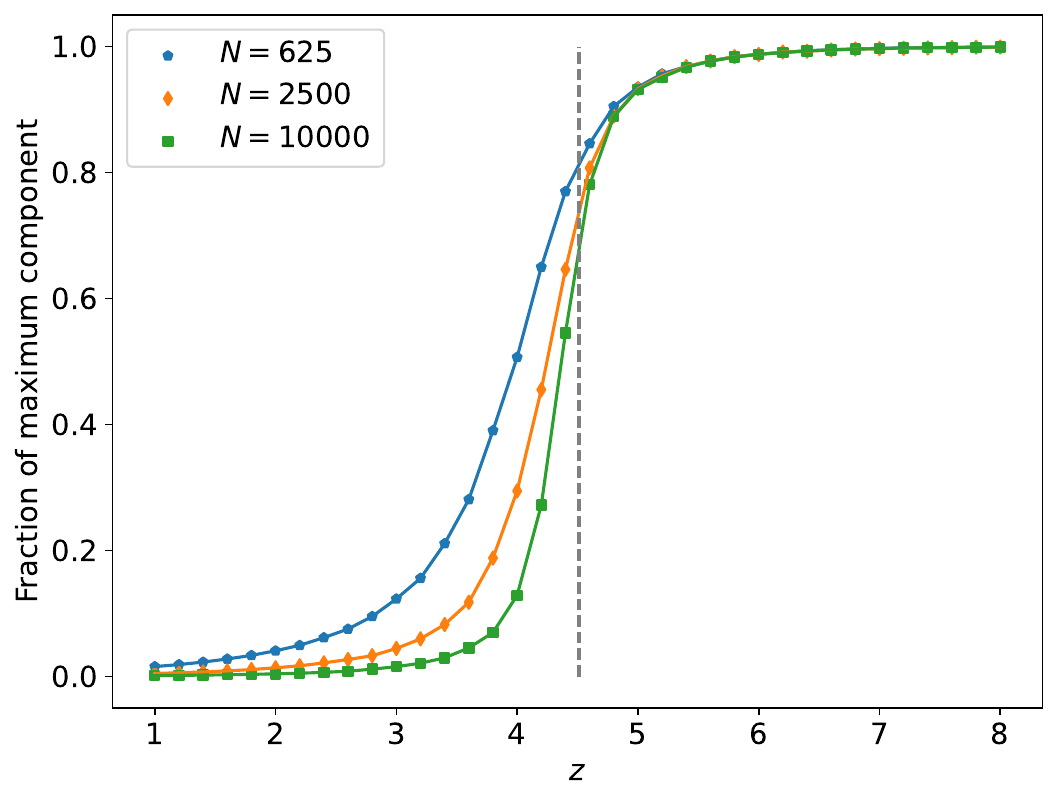}
    \end{subfigure}
    \caption{Left: Component size distribution of systems of size $N=10000$ and different $z$ (1000 realisations). The dashed line indicates a power law with the Fisher exponent $187/91$. The deviation of the points from 1 at the system size although there is only a single component for $z=16,32$ is due to the normalization. Right: The average of all maximum components of an ensemble of 1000 for $N=625$ and 100 for $N=2500,10000$ dependent on $z$. With increasing system size the curve becomes steeper in proximity of the critical connectivity radius $4.51218\leq z_c \leq 4.51228$~\cite{balisterContinuumPercolationSteps2005} (grey dashed line).}
    \label{fig:component_sizes}
\end{figure}

Depending on the mean degree, the graph has a system-spanning component or not.
At a critical $z_c$ (percolation threshold) there is a phase transition from a graph with only finite components to a graph that contains an infinite component in the thermodynamic limit~\cite{stauffer1992introduction,quintanillaEfficientMeasurementPercolation2000}.
This is illustrated in Fig.~\ref{fig:component_sizes}.
Apparently, the threshold lies at approximately $z=4.5$ which is in accordance with the previously reported result $z=4.512$ \cite{balbergExcludedVolumeIts1984,alonNewHeuristicPercolation1991,pikePercolationConductivityComputer1974a, balisterContinuumPercolationSteps2005,quintanillaEfficientMeasurementPercolation2000}. At the percolation threshold the component size distribution must asymptotically follow a power law with the Fisher exponent of  $-187/91$~\cite{stauffer1992introduction}. 
With increasing $z$, the proportion of sites in the system-spanning component increases, and the average number and size of small components decreases, leading to a clear gap in the data. For $z=16$, our data in Fig.~\ref{fig:component_sizes} show less than 1 percent of  small components, and all of them have size 1; for $z=32$ small components are so rare that they do not occur in the studied ensemble of graphs. 

\section{Methods}
For measuring the localisation of a state $n$, we compute its participation ratio
\begin{align}
    P_n \coloneqq \frac{\left(\sum_{i=1}^N\psi_{n,i}^2\right)^2}{\sum_{i=1}^N\psi_{n,i}^4}~.
\end{align}
This takes the value 1 for a state which is fully localised on a single node.
It is equal to $N$ for an eigenmode which is uniformly distributed over all nodes, and it approaches $(2/3)^2$ for a sinusoidal state on a square lattice with a wavelength much larger than the lattice constant.
The states themselves and their corresponding eigenvalues are obtained via numerically diagonalizing the graph matrices with the function \texttt{eigen()} in the Julia (Version 1.10.1) library \texttt{Linear Algebra}.

\section{Eigenvectors}
\begin{figure}[H]
    \centering
    \begin{subfigure}[b]{0.24\textwidth}
        \includegraphics[width=\textwidth]{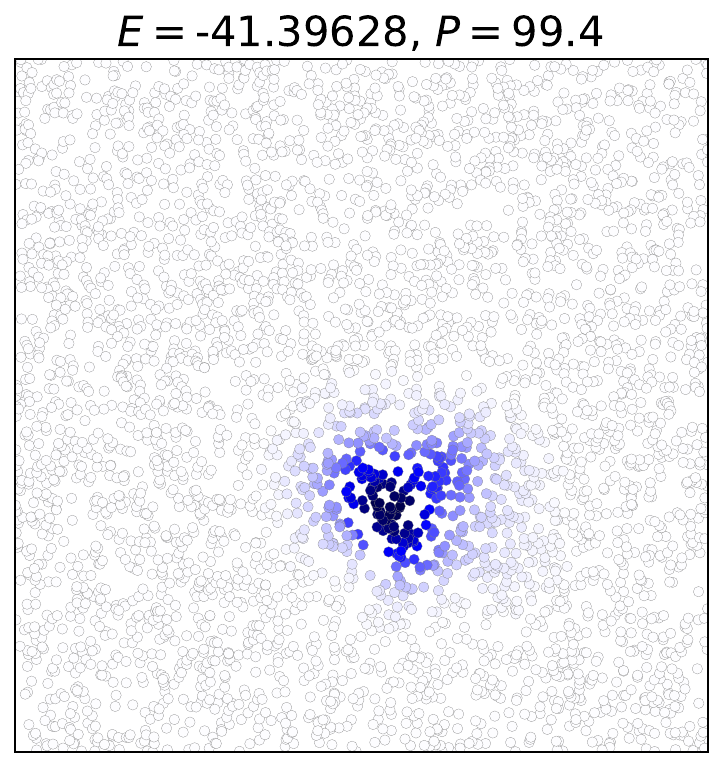}
    \end{subfigure}
    \hfill
    \begin{subfigure}[b]{0.24\textwidth}
        \includegraphics[width=\textwidth]{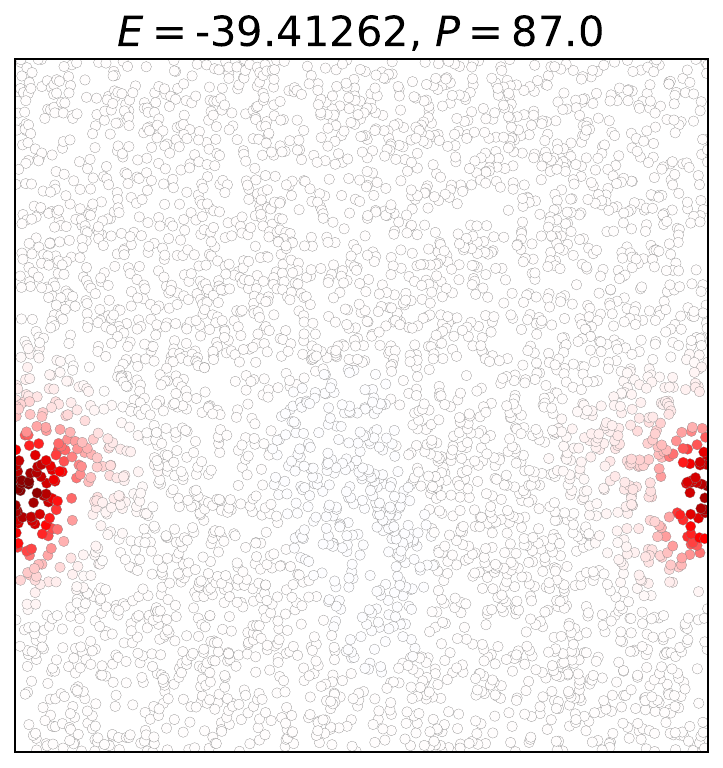}
    \end{subfigure}
    \hfill
    \begin{subfigure}[b]{0.24\textwidth}
        \includegraphics[width=\textwidth]{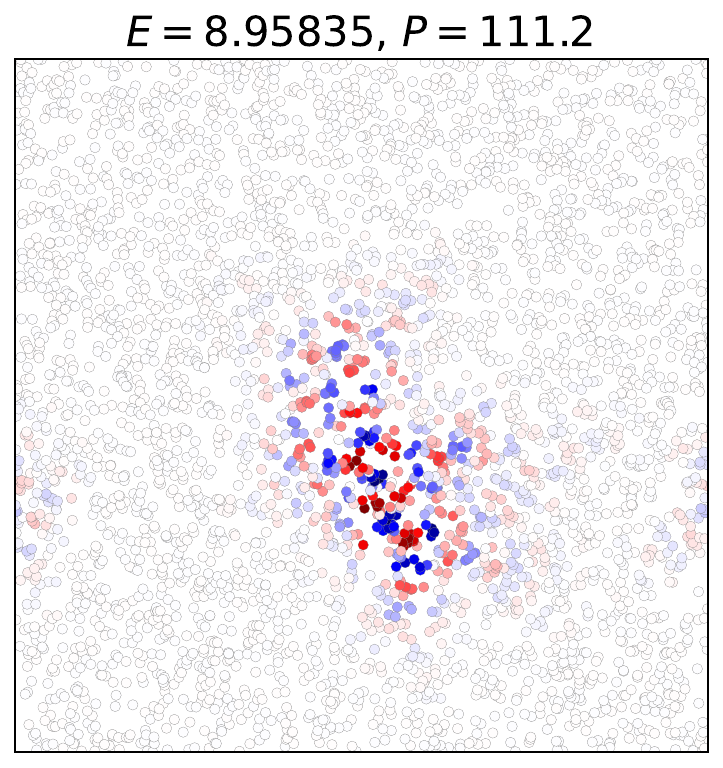}
    \end{subfigure}
    \hfill
    \begin{subfigure}[b]{0.24\textwidth}
        \includegraphics[width=\textwidth]{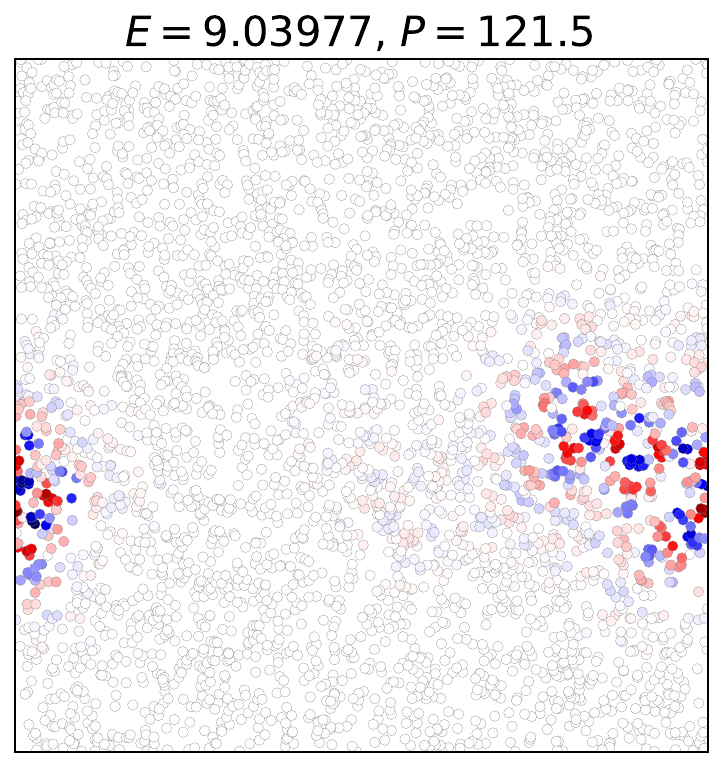}
    \end{subfigure}
    \hfill
    \begin{subfigure}[b]{0.24\textwidth}
        \includegraphics[width=\textwidth]{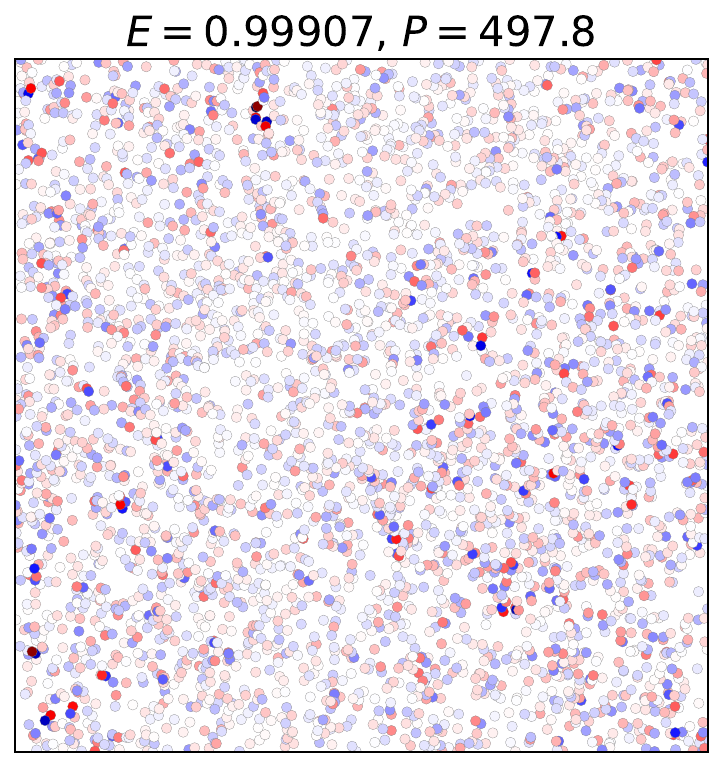}
    \end{subfigure}
    \hfill
    \begin{subfigure}[b]{0.24\textwidth}
        \includegraphics[width=\textwidth]{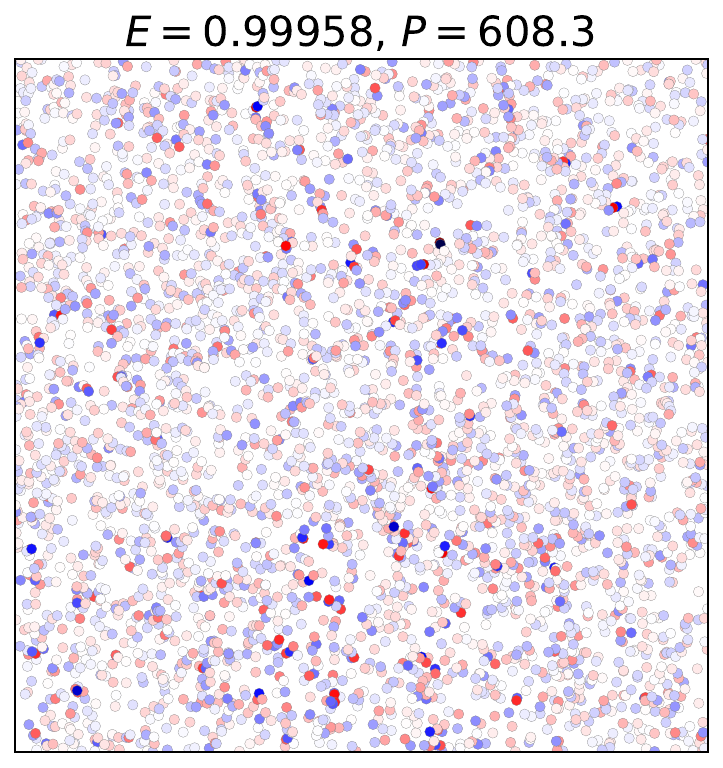}
    \end{subfigure}
    \hfill
    \begin{subfigure}[b]{0.24\textwidth}
        \includegraphics[width=\textwidth]{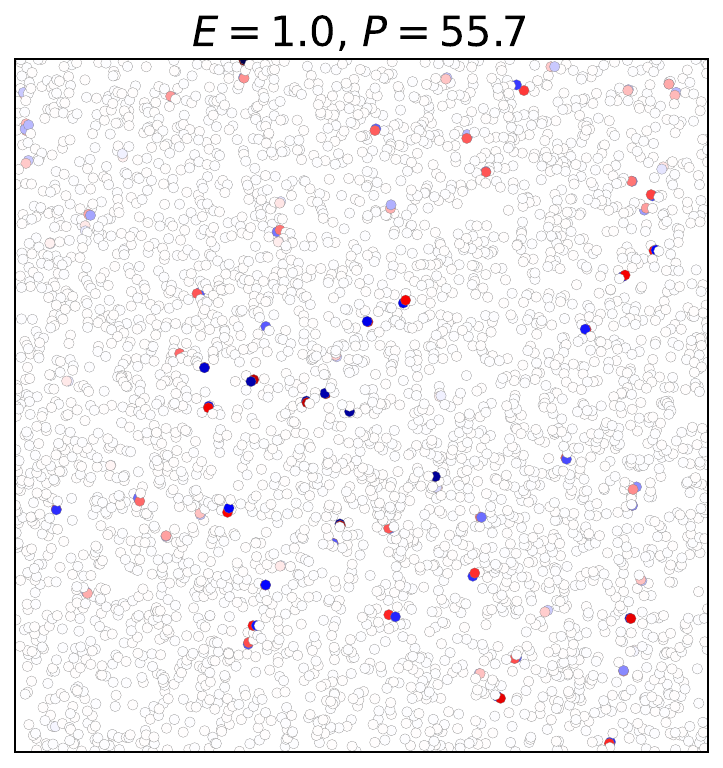}
    \end{subfigure}
    \hfill
    \begin{subfigure}[b]{0.24\textwidth}
        \includegraphics[width=\textwidth]{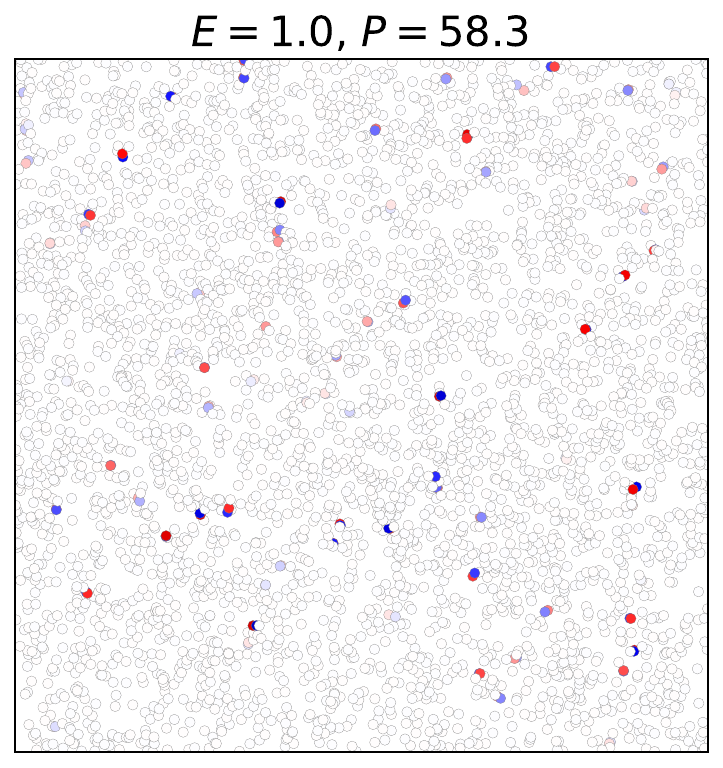}
    \end{subfigure}
    \caption{In the top row, the first two eigenvectors associated with the eigenvalues at the band edges of the AM (($N=4000,~z=32$) are shown. Red indicates a high, blue a low, and white an amplitude of zero.
    For negative (positive) eigenvalues, neighbouring entries of the eigenvector tend to have the same (the opposite) sign. 
    In the bottom row, eigenvectors close to $E=1$ and at $E=1$ are shown.
    At $E=1$, the eigenvector is fully localised on the orbits, while for $E$ close to 1 the participation ratio is large.}
    \label{fig:eigenvectors_AM}
\end{figure}

Before diving into the results, it is insightful to have a glance at the eigenvectors of the systems.
In the AM, the spectrum comprises positive and negative eigenvalues.
The eigenvectors associated with the most negative and the largest positive eigenvalue are localised in a small spatial region.
For negative energies, eigenvector entries of neighbouring sites have preferentially the same sign, whereas for positive energies, neighbouring entries tend to have the opposite sign (see top row of Fig.~\ref{fig:eigenvectors_AM}). The eigenvectors with the largest participation ratio are associated with positive energies close to 1 (see left two pictures in the bottom row of Fig.~\ref{fig:eigenvectors_AM}), and the correlations of the eigenvalue entries appear very short-ranged. At $E=1$, the eigenvectors are fully localised on orbits. Due to the degeneracy of this eigenvalue, the diagonalization algorithm produces linear combinations of these fully localised eigenvectors and thus the participation ratio is much larger than the expected value 2 (see last two pictures of Fig.~\ref{fig:eigenvectors_AM}). 

\begin{figure}[H]
    \centering
    \begin{subfigure}[b]{0.24\textwidth}
        \includegraphics[width=\textwidth]{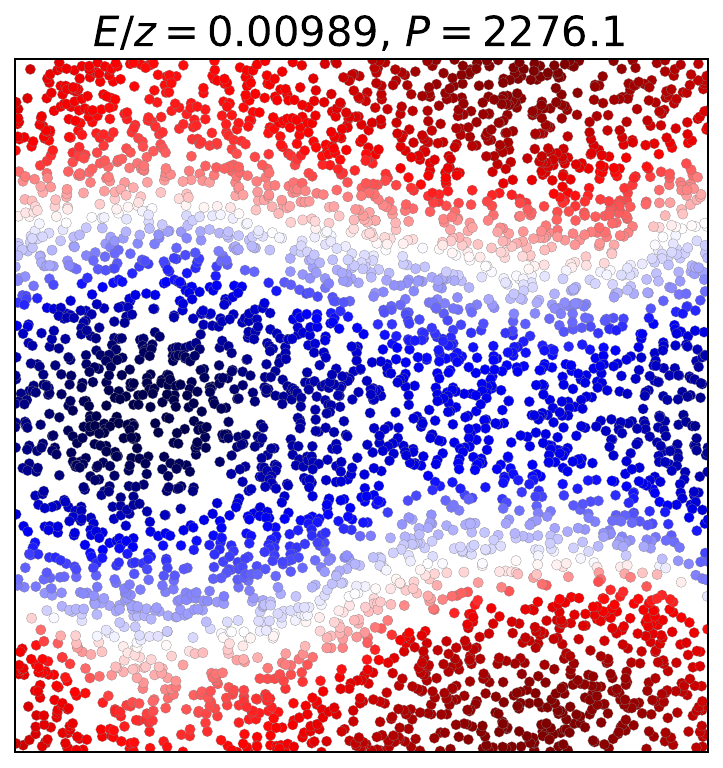}
    \end{subfigure}
    \hfill
    \begin{subfigure}[b]{0.24\textwidth}
        \includegraphics[width=\textwidth]{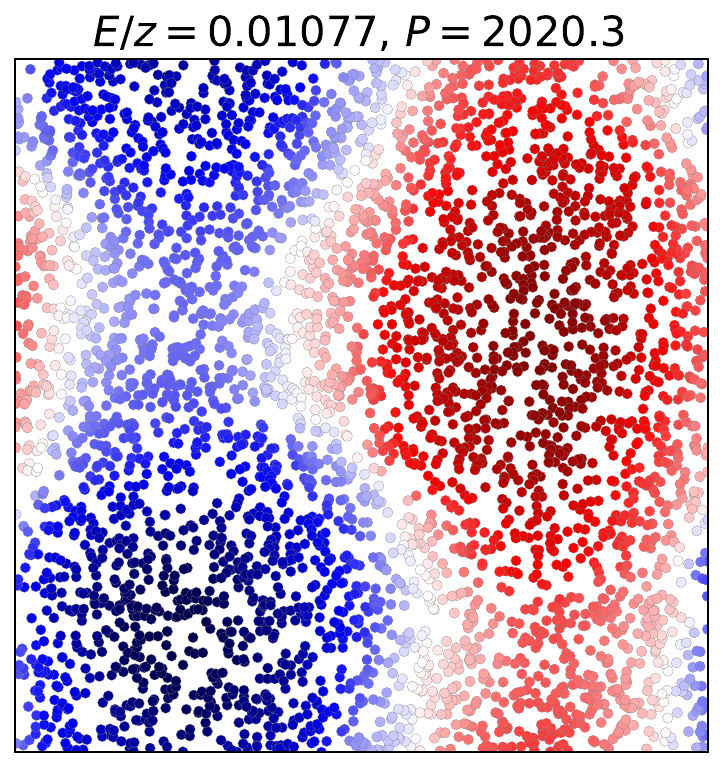}
    \end{subfigure}
    \hfill
    \begin{subfigure}[b]{0.24\textwidth}
        \includegraphics[width=\textwidth]{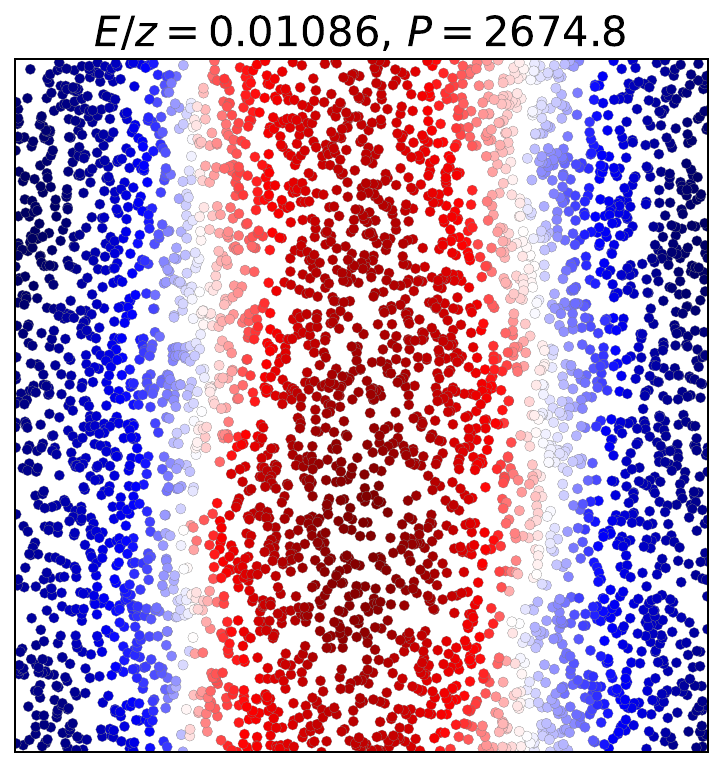}
    \end{subfigure}
    \hfill
    \begin{subfigure}[b]{0.24\textwidth}
        \includegraphics[width=\textwidth]{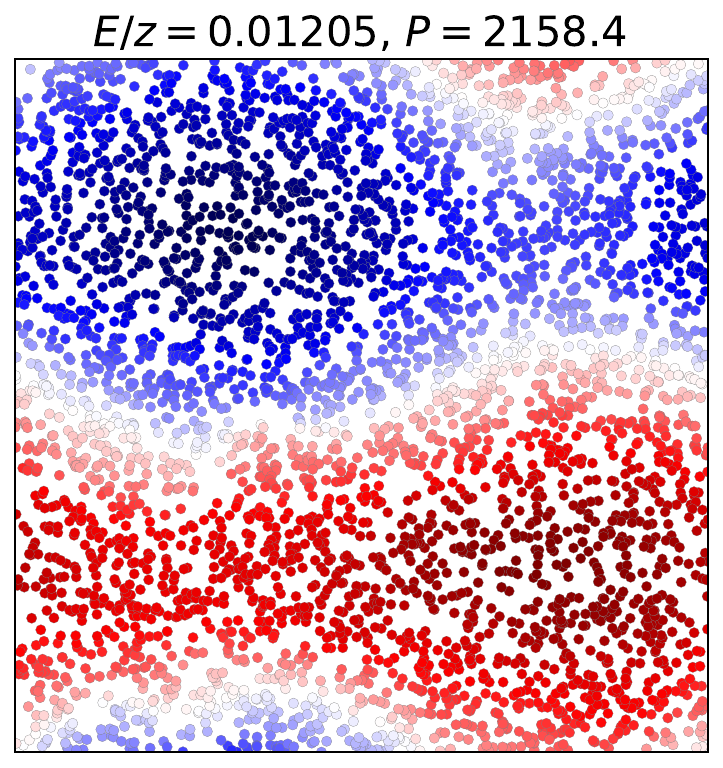}
    \end{subfigure}
    \hfill
    \begin{subfigure}[b]{0.24\textwidth}
        \includegraphics[width=\textwidth]{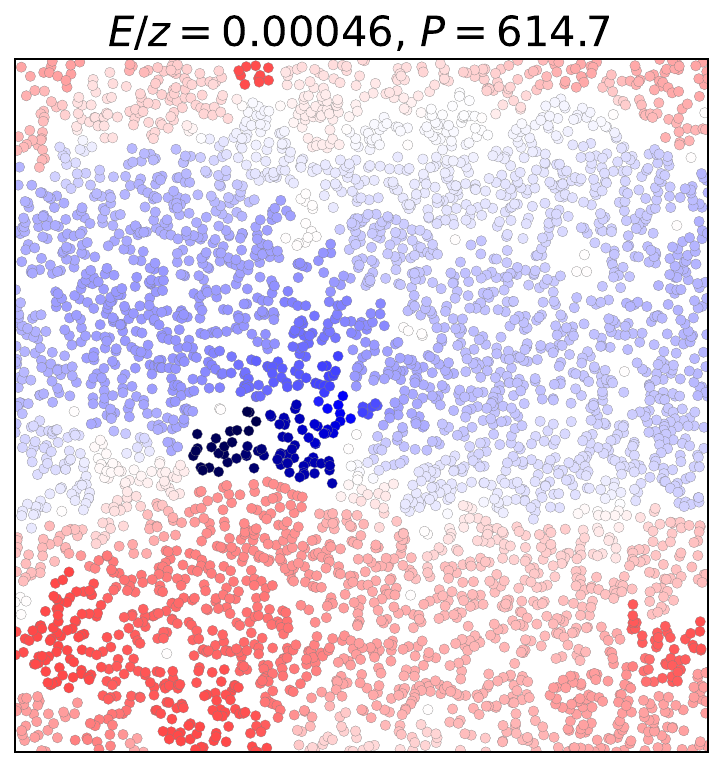}
    \end{subfigure}
    \hfill
    \begin{subfigure}[b]{0.24\textwidth}
        \includegraphics[width=\textwidth]{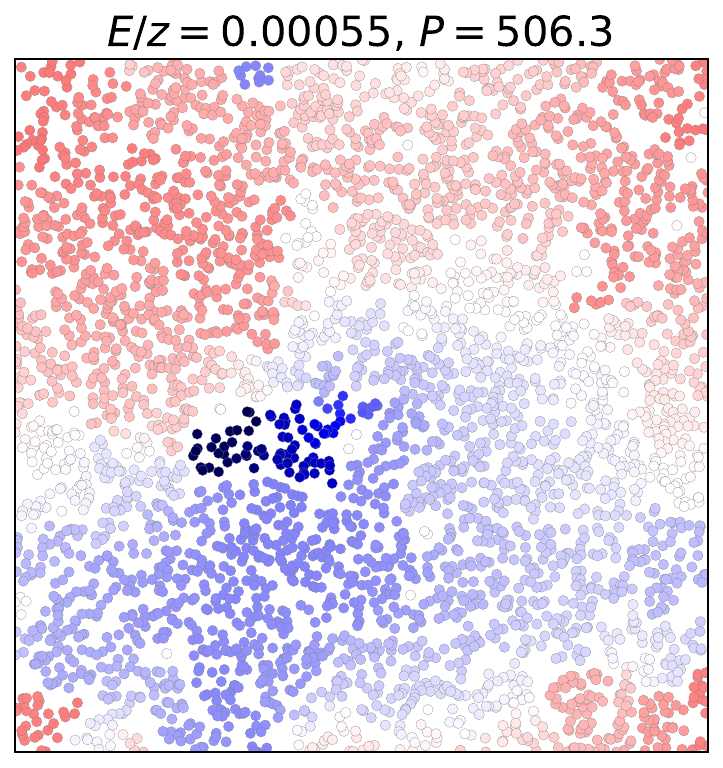}
    \end{subfigure}
    \hfill
    \begin{subfigure}[b]{0.24\textwidth}
        \includegraphics[width=\textwidth]{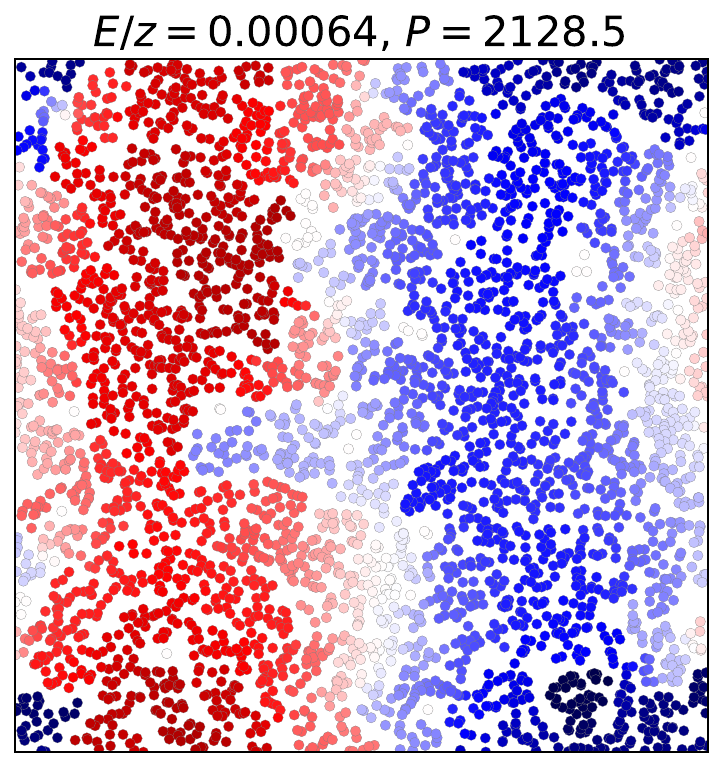}
    \end{subfigure}
    \hfill
    \begin{subfigure}[b]{0.24\textwidth}
        \includegraphics[width=\textwidth]{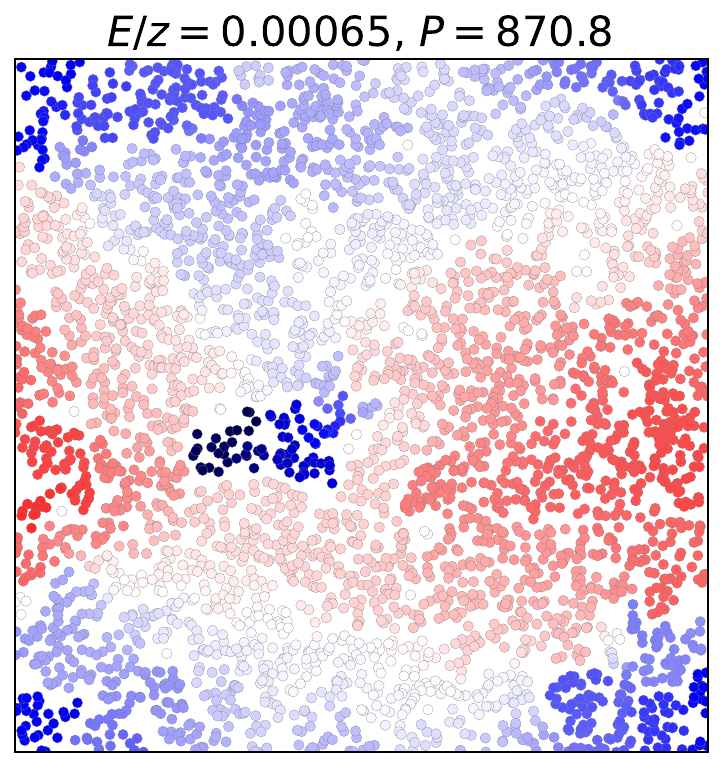}
    \end{subfigure}
    \caption{Eigenvectors of the four smallest eigenvalues greater than zero for the LM. Top row: $N=4000,~z=32$; bottom row: $N=4000,~z=6$.
    Red indicates positive, blue negative, and white zero amplitude. In both cases the sinusoidal form of the eigenmodes is apparent but for $z=6$ weakly-connected substructures are visible.}
    \label{fig:eigenvectors_LM}
\end{figure}

In the LM, the eigenvectors of the four smallest eigenvalues larger than zero are very similar to the sine functions of an ordered lattice (Fig.~\ref{fig:eigenvectors_LM}).
In contrast to the ordered system, the degeneracy of degree four is lifted, and the four energies differ slightly. For $z=8$, disorder is stronger than for $z=32$, and weakly connected subsets of nodes have an amplitude that deviates considerably from their surroundings.

\section{Density of states}

\begin{figure}[H]
    \centering
    \begin{subfigure}[b]{0.49\textwidth}
        \includegraphics[width=\textwidth]{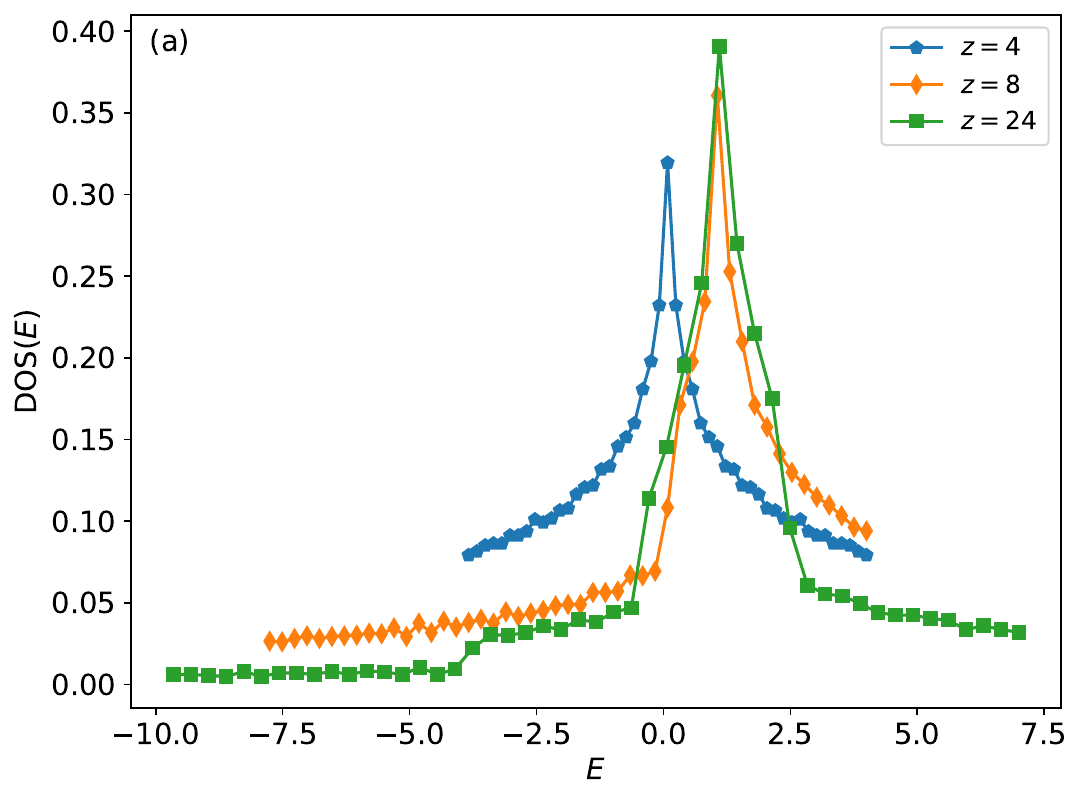}
    \end{subfigure}
\hfill
    \begin{subfigure}[b]{0.49\textwidth}
        \includegraphics[width=\textwidth]{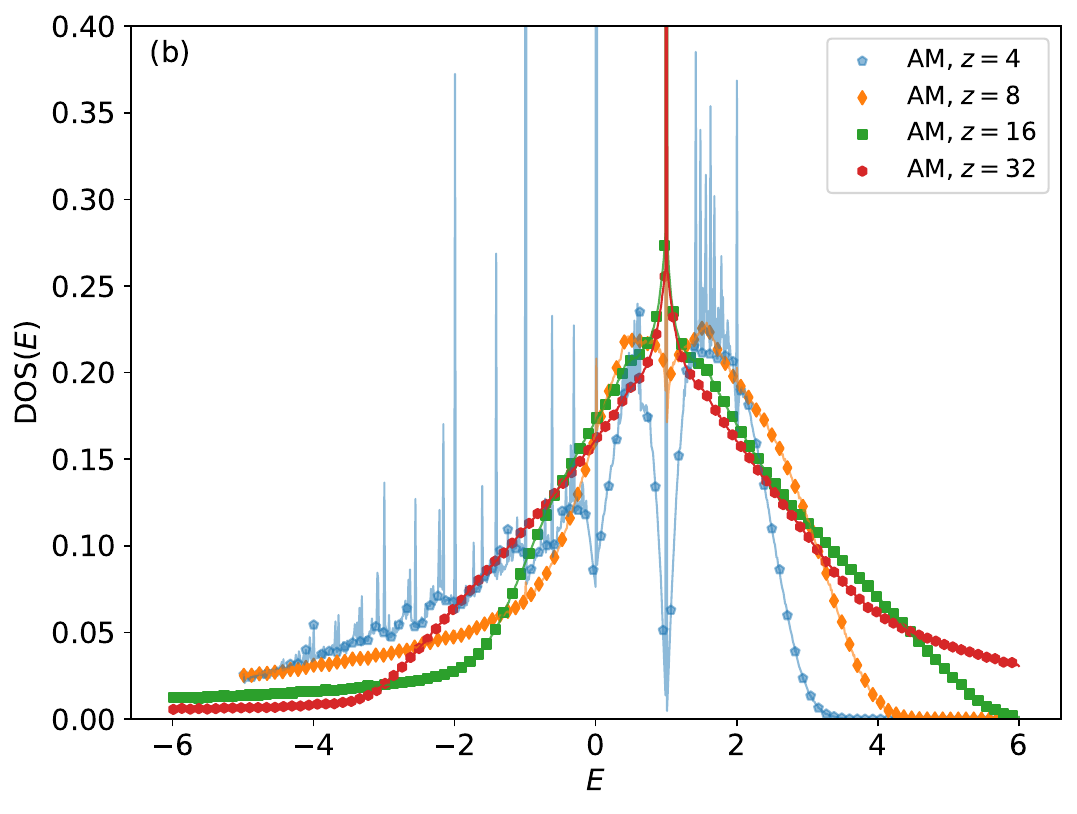}
    \end{subfigure}
\hfill
\begin{subfigure}[b]{0.49\textwidth}
        \includegraphics[width=\textwidth]{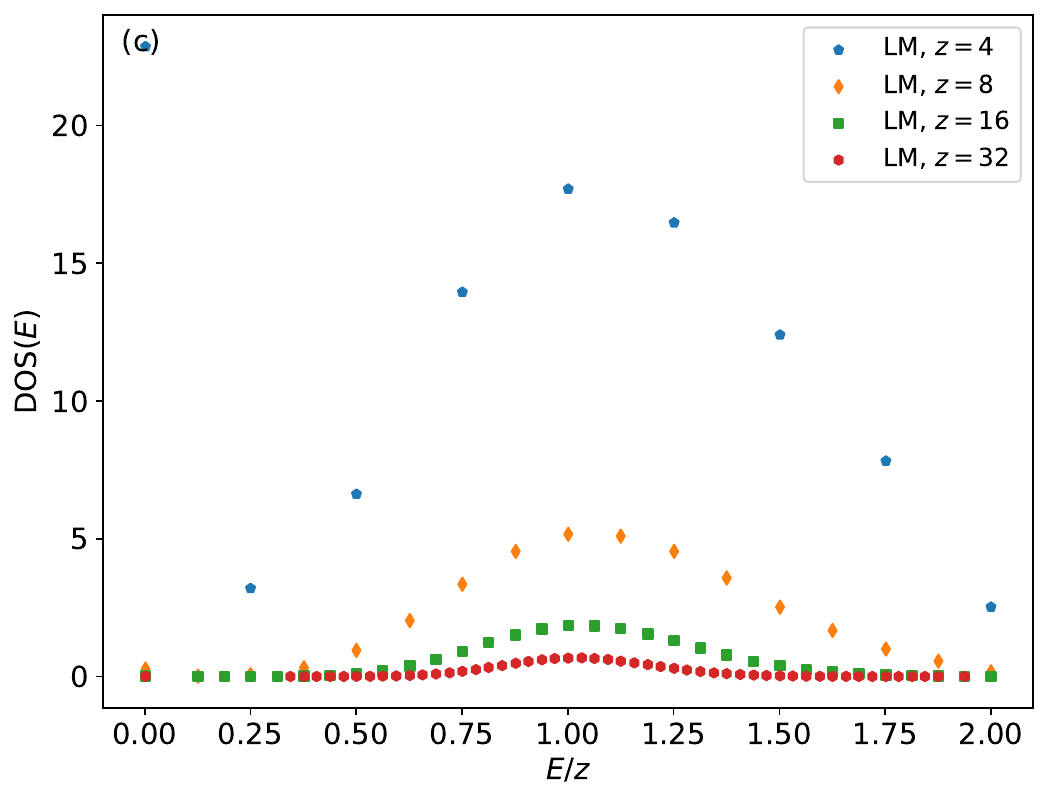}
    \end{subfigure}
\hfill
    \begin{subfigure}[b]{0.49\textwidth}
        \includegraphics[width=\textwidth]{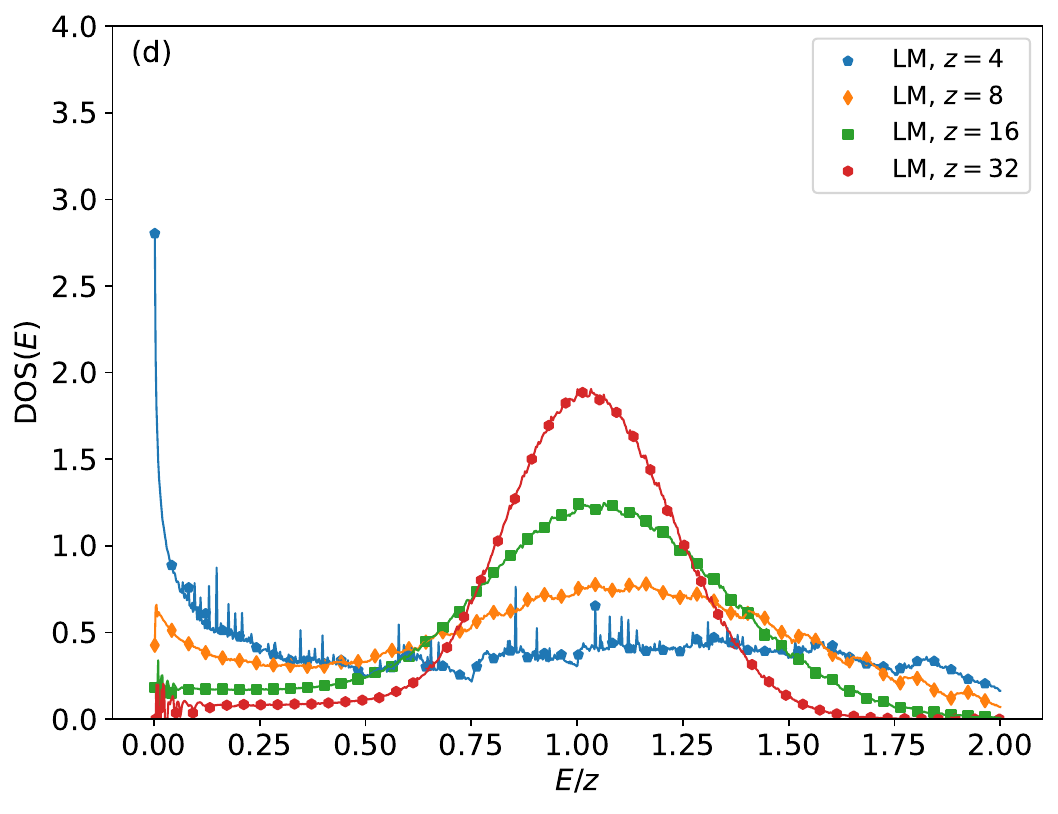}
    \end{subfigure}
    \caption{(a) Density of states (DOS) of the (negative) adjacency matrix of a regular lattice with 4, 8, and 24 neighbours. The DOS of the regular 2D lattice ($z=4$) has been calculated analytically in~\cite{economou2006}, showing that there is a logarithmic divergence in the centre of the band and the band itself extends over the range of $8$. (b) DOS of the AM on an RGG (c) Discrete spectrum of the LM. (d) Continuous spectrum of the LM. The last three graphs were evaluated for an ensemble of 1000 systems and $N=10^4$.}
    \label{fig:DOS}
\end{figure}

The density of states of the AM has a peak around $E=1$ with a discrete maximum at exactly $E=1$ due to the orbits, see Fig.~\ref{fig:DOS}(b). 
The maximum becomes more symmetric for increasing $z$ and approaches a triangular shape for $z=32$. For $z=4$ and $z=8$, there is a funnel around $E=1$, which must be due to the large number of orbits for these $z$ values, which deplete eigenvalues in the vicinity  of 1. The tails extend beyond the band edges of the ordered system (see Fig.~\ref{fig:DOS}(a)), with the decrease being slower for negative than for positive $E$. 

The spectrum of the LM is divided into a discrete part with integer $E$ values due to the orbits (Fig.~\ref{fig:DOS}(c)) and a continuous part (Fig.~\ref{fig:DOS}(e)), both of which have a maximum around $E/z=1$.
For $z=4$ and $z=8$, where the graph contains many small components, the density of states to the left of the peak rises with decreasing energy because small components cannot have large eigenvalues. This leads to an accumulation of smaller eigenvalues. 
In particular for $z=32$, the density of states shows pronounced oscillations for the smallest values of $E$.
These stem from nearly periodic eigenfunctions that resemble those of the ordered system, see Fig.~\ref{fig:eigenvectors_LM}, and that have an integer number of wavelengths. By comparing to the eigenmodes in this energy range we could see  that the left-most peak belongs to the (1,1)-modes, the second one to linear combinations of the (2,1)-modes, and so on. For smaller $z$, these eigenmodes become less regular and their energies spread over larger intervals, so that these oscillations become less visible.

At intermediate energies, there are also oscillations, at least for $z=4$ and $z=8$. The minima of these oscillations are at integer values of $E$, indicating a correlation with the discrete part of the spectrum, similar to the AM, where there is a funnel around $E=1$ because the integer-energy orbits deplete neighbouring energy values.

The height of the maximum of the discrete part of the spectrum decreases superlinearly with $z$. 
This is in accordance with the findings of Dettmann and Knight~\cite{dettmannSymmetricMotifsRandom2018} who found that the average number of orbits per site goes to zero in the limit $z\to \infty$ in two dimensions.

\section{Participation ratio}
\subsection{Probability distribution of the participation ratio}

\begin{figure}[H]
    \centering
\begin{subfigure}[b]{0.49\textwidth}
        \includegraphics[width=\textwidth]{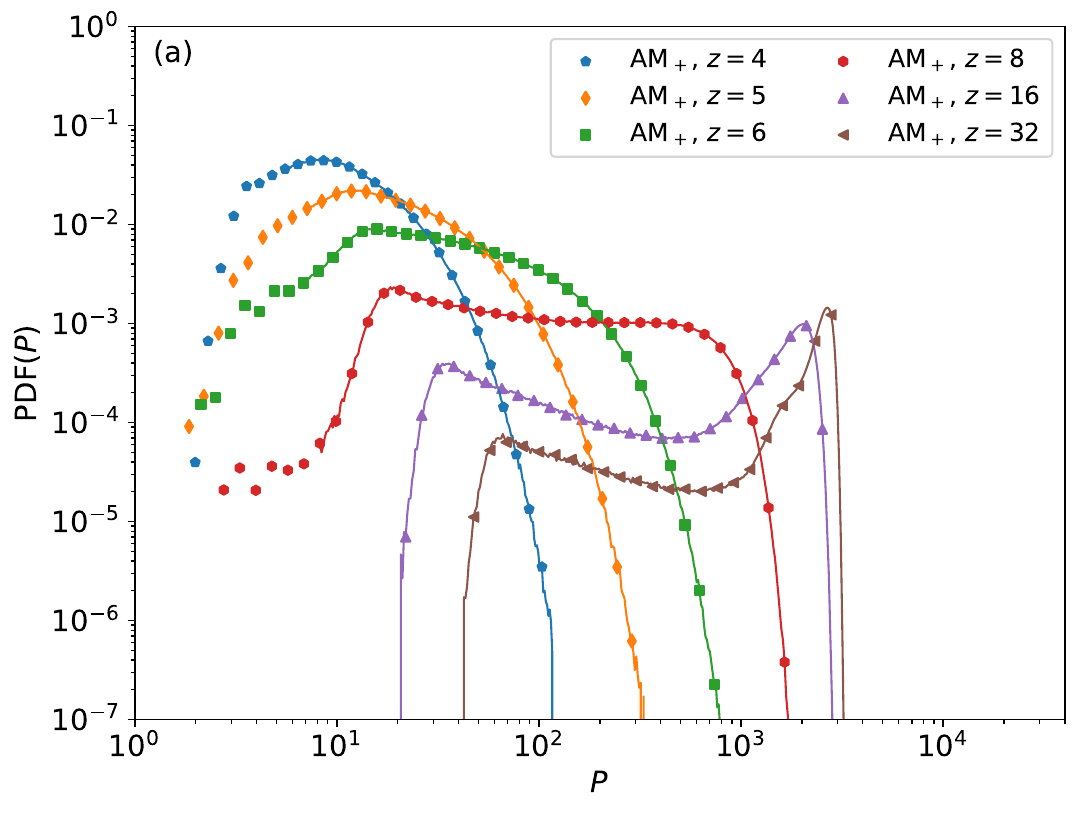}
    \end{subfigure}
\hfill
    \begin{subfigure}[b]{0.49\textwidth}
        \includegraphics[width=\textwidth]{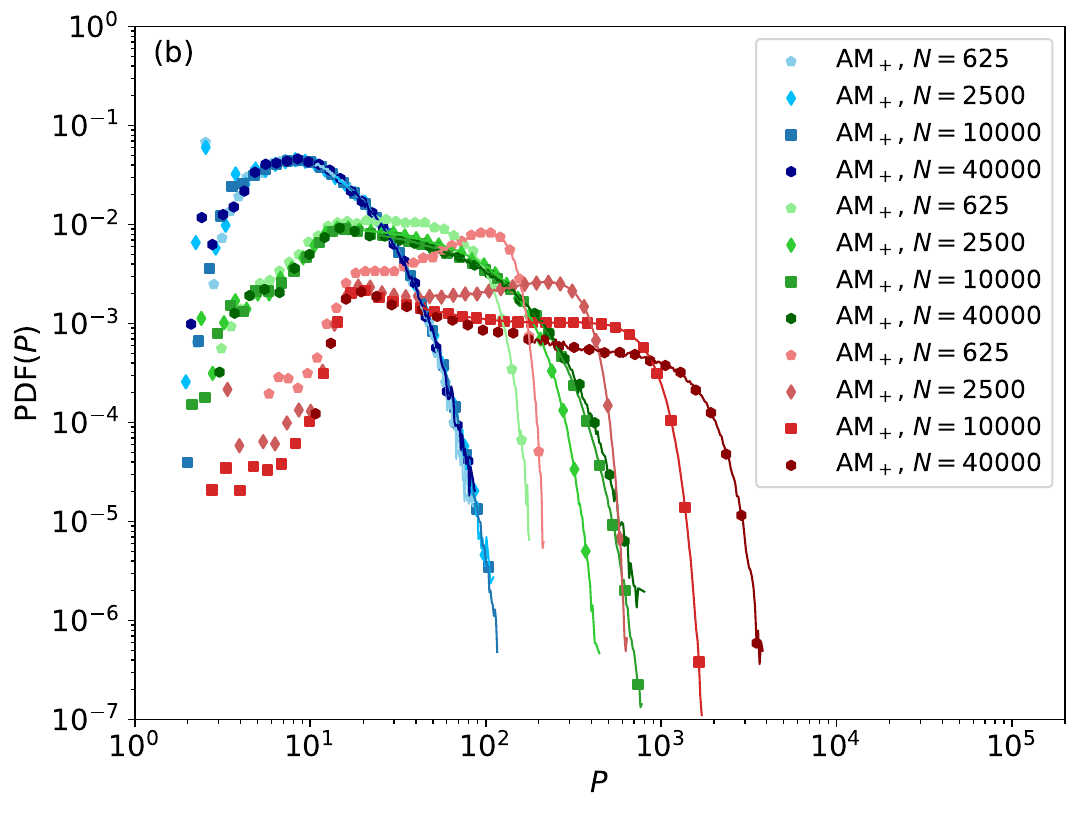}
    \end{subfigure}
    \hfill
\begin{subfigure}[b]{0.49\textwidth}
        \includegraphics[width=\textwidth]{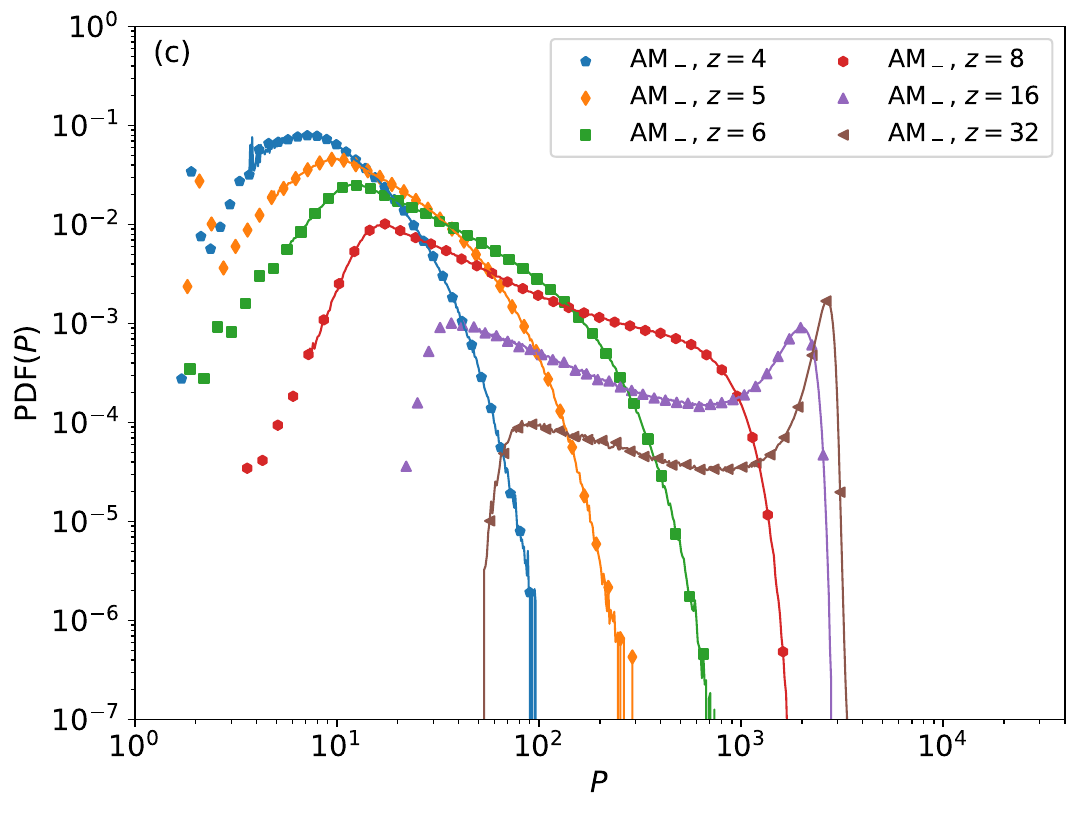}
    \end{subfigure}
\hfill
    \begin{subfigure}[b]{0.49\textwidth}
        \includegraphics[width=\textwidth]{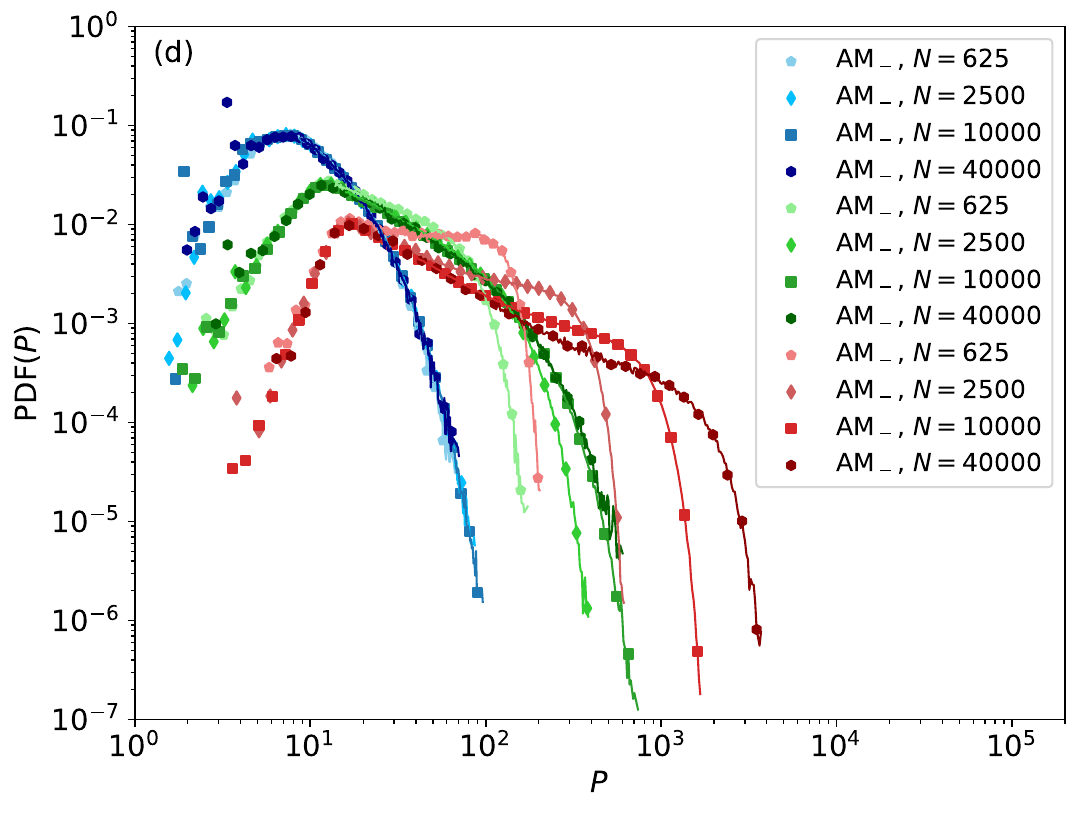}
    \end{subfigure}
    \hfill
    \begin{subfigure}[b]{0.49\textwidth}
        \includegraphics[width=\textwidth]{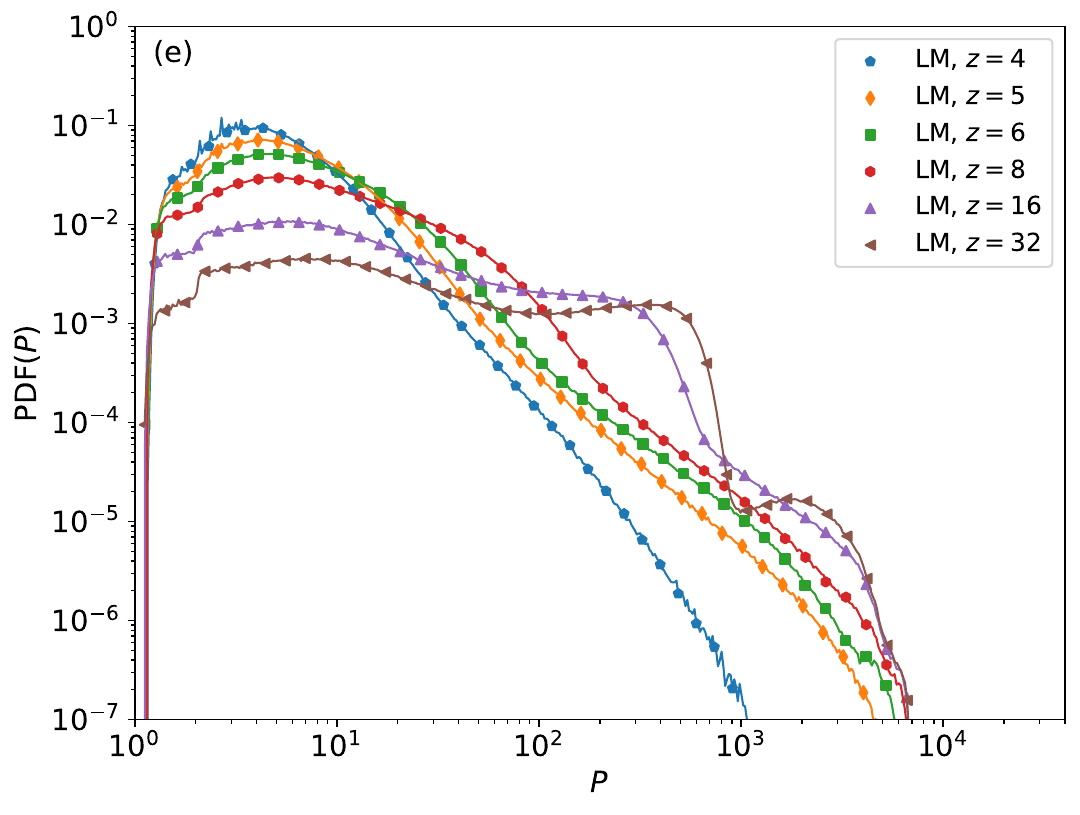}
    \end{subfigure}
\hfill
    \begin{subfigure}[b]{0.49\textwidth}
        \includegraphics[width=\textwidth]{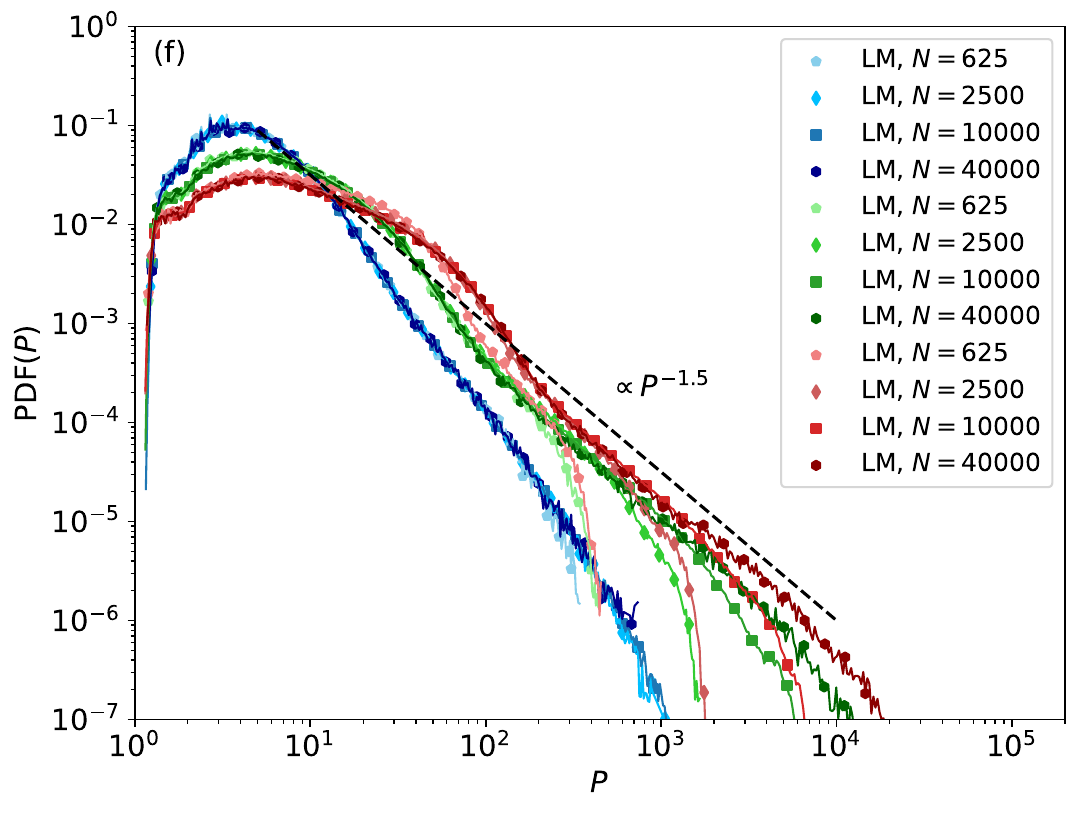}
    \end{subfigure}
    \caption{Distribution of the participation ratio for the continuous part of the spectrum. On the left-hand side $z$ is varied for constant $N=10^4$. On the right-hand side $z=4$ is plotted in  blue, $z=6$ in green, and $z=8$ in red for different system sizes. (a)-(d) show the AM with participation ratios belonging to positive eigenvalues in the top row and those arising from eigenvectors with negative eigenvalues in the second row. (e) and (f) show the LM, with the asymptotically expected power law of exponent $-1.5$ indicated by a dashed line in (f). The ensemble size is $10^3$ apart from $N=4\times 10^4$ where it is $10$.}
    \label{fig:PPDF}
\end{figure}
 
The probability distribution of the participation ratio (PPDF) becomes broader with increasing $z$, and its mean value moves to the right, see Fig.~\ref{fig:PPDF}. 
With increasing $z$,  finite-size effects become stronger, showing peaks at large values of $P$. 
Even for the AM, where we expect complete localisation of all modes, there are considerable finite-size effects for $z\ge 8$, indicating that the eigenmodes extend to the system boundaries, see Fig.~\ref{fig:eigenvectors_AM}. 
This is true for positive as well as negative energy eigenvalues (first and second row of the figure, respectively). 
For $z=8$, it appears as if the maximum participation ratio of the AM increases approximately with a power law, as indicated by the approximately equidistant cutoffs of the red data in Fig.~\ref{fig:PPDF}(b) and (d), but the exponent of this apparent power law decreases with decreasing $z$ and increasing system size, where finite-size effects are expected to become weaker. 
Fig.~\ref{fig:PPDF}(b) and (d) show that for $z=6$ (green data) the curves for $N=10^4$ and $4\times10^4$ almost coincide.

The PPDF of the LM shows a long tail for large $P$ for $z>4$. Due to the conservation law in the LM, there must be system-spanning modes and therefore large values of $P$ unless $z$ is so small that there are no system-spanning graph components. 
For $z=16$ and 32, a hump emerges at $P\approx 5\times10^2$ followed by a sharp drop (Fig.~\ref{fig:PPDF}(e)). This drop means that the number of eigenmodes decreases drastically, and we ascribe it to the strong decrease of the DOS to the left of the peak at $E/z=1$ in Fig.~\ref{fig:DOS}(d).  
The peak at the right end of the PPDF for $z=32$ is due to finite-size effects, and so are the cutoffs of all curves for $z>4$. For sufficiently large system sizes, we expect a power law for larger values of $P$, where eigenmodes are still localised but system size is large enough to show modes that are reminiscent of the periodic eigenmodes of a fully ordered system. Due to the randomness of the system, the amplitudes of neighbouring maxima of these modes differ by a random amount of the order $\sqrt{\lambda}$ (with $\lambda$ being the wavelength of the periodic mode), and perform a random walk in both system dimensions.

Based on this idea, we can derive the exponent of the power law, in a similar way as we did in \cite{schaefer2025scaling}. 
In each dimension, the random walk extends over $\lambda$ wavelengths, meaning such an eigenmode is extended over $\propto\lambda^2\propto 1/E$ nodes in each dimension,  which results in $P\propto 1/E^2$ in two dimensions. With this information, the PPDF can be obtained from the DOS of the energy, which approaches in two dimensions a constant for sufficiently small $E$. This implies
\begin{equation}\mathrm{PPDF}(P) = \mathrm{DOS}(E)\frac{ \mathrm{d} P}{\mathrm{d}E} \propto 1/E^3 \propto P^{-1.5}\, .
\label{powerlaw}
\end{equation}
We see that the data in Fig.~\ref{fig:PPDF}(f) approach this power law for the larger system sizes.

\subsection{Dependence of the participation ratio on the energy}
\begin{figure}[H]
    \centering
    \begin{subfigure}[b]{0.49\textwidth}
        \includegraphics[width=\textwidth]{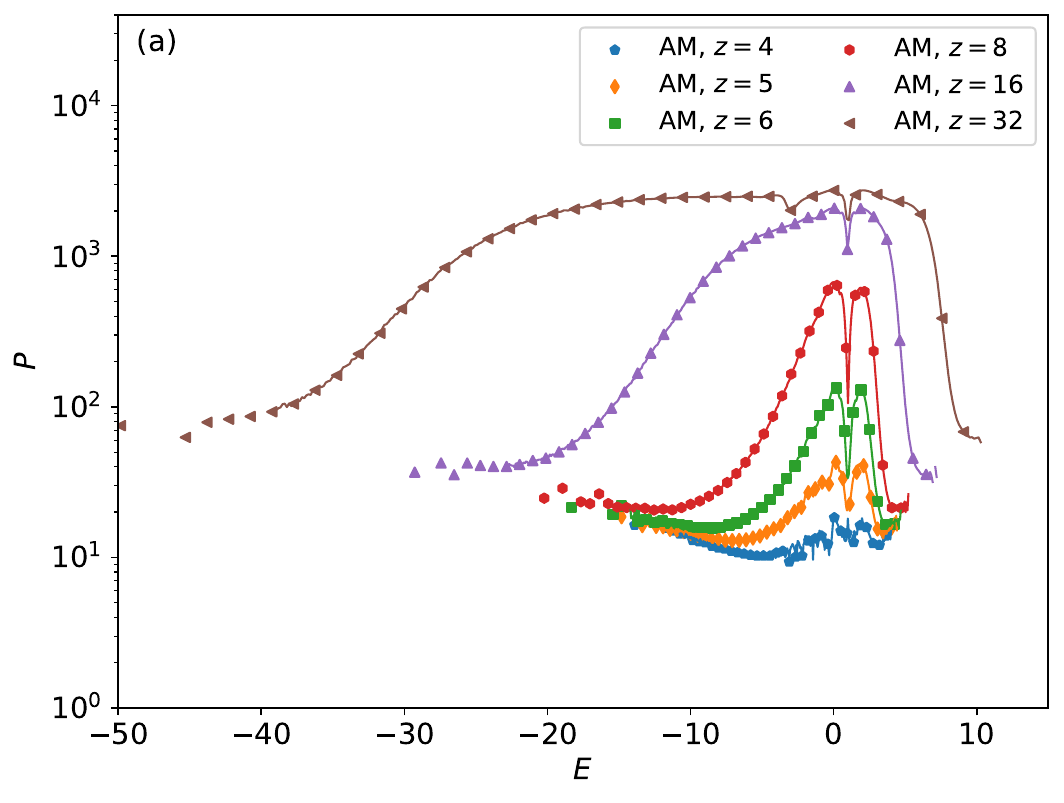}
    \end{subfigure}
\hfill
    \begin{subfigure}[b]{0.49\textwidth}
        \includegraphics[width=\textwidth]{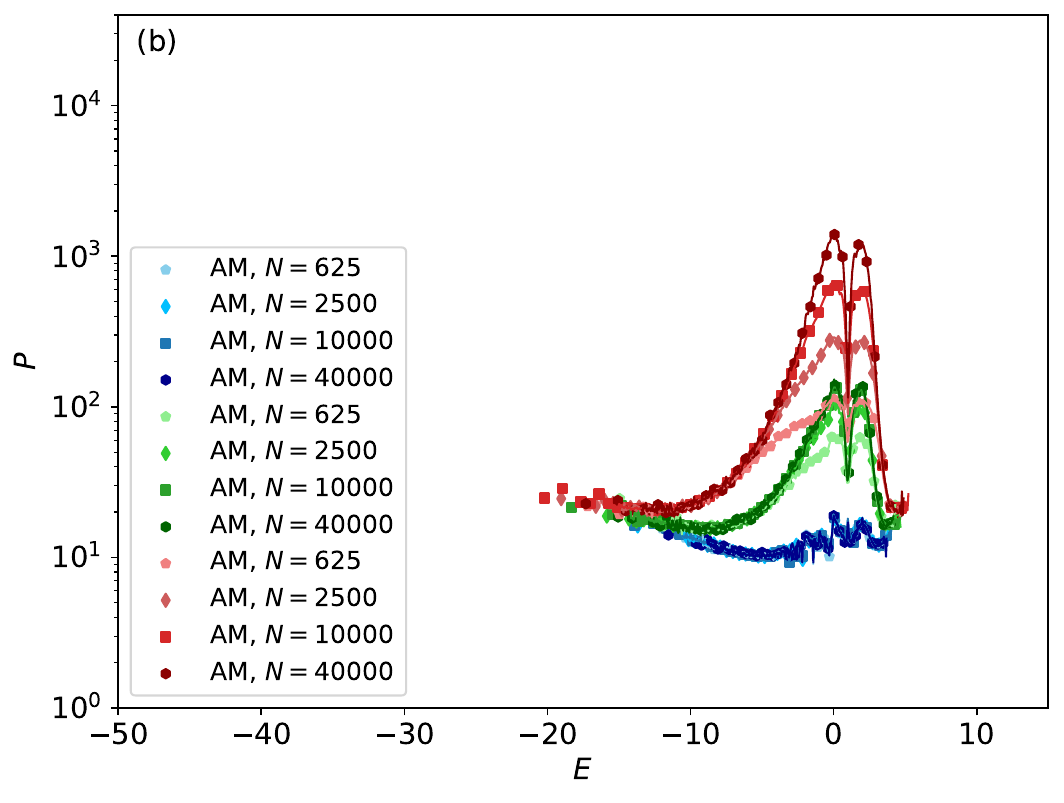}
    \end{subfigure}
    \hfill
    \begin{subfigure}[b]{0.49\textwidth}
        \includegraphics[width=\textwidth]{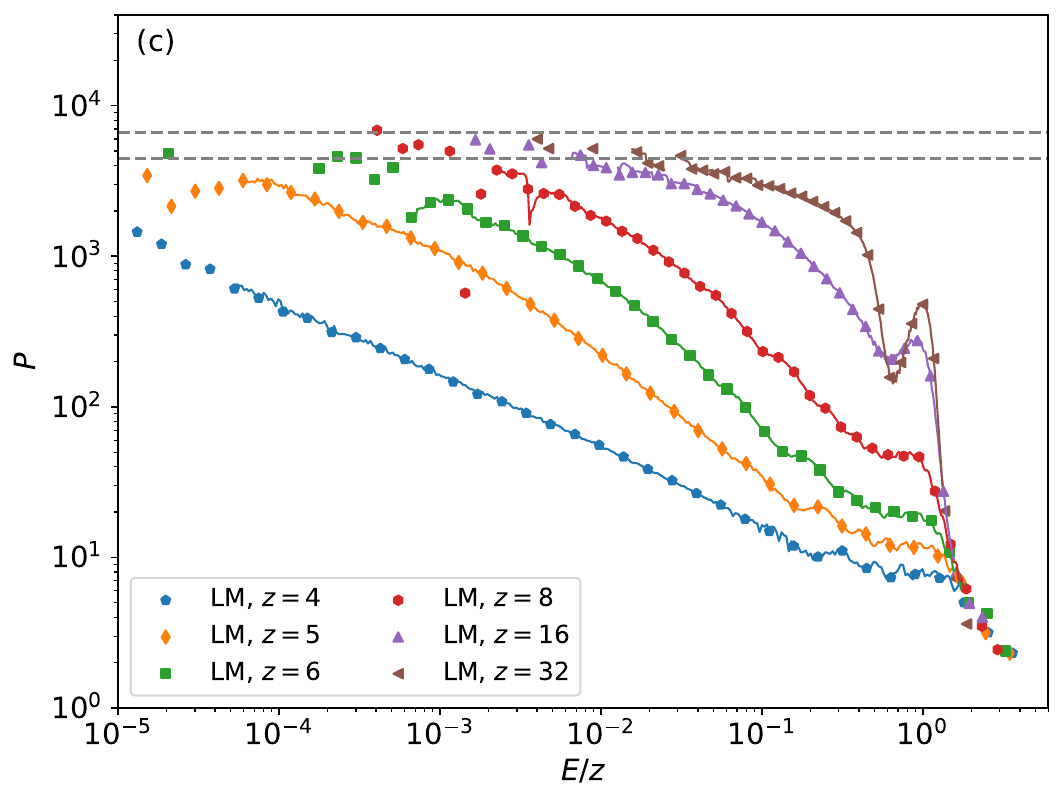}
    \end{subfigure}
\hfill
    \begin{subfigure}[b]{0.49\textwidth}
        \includegraphics[width=\textwidth]{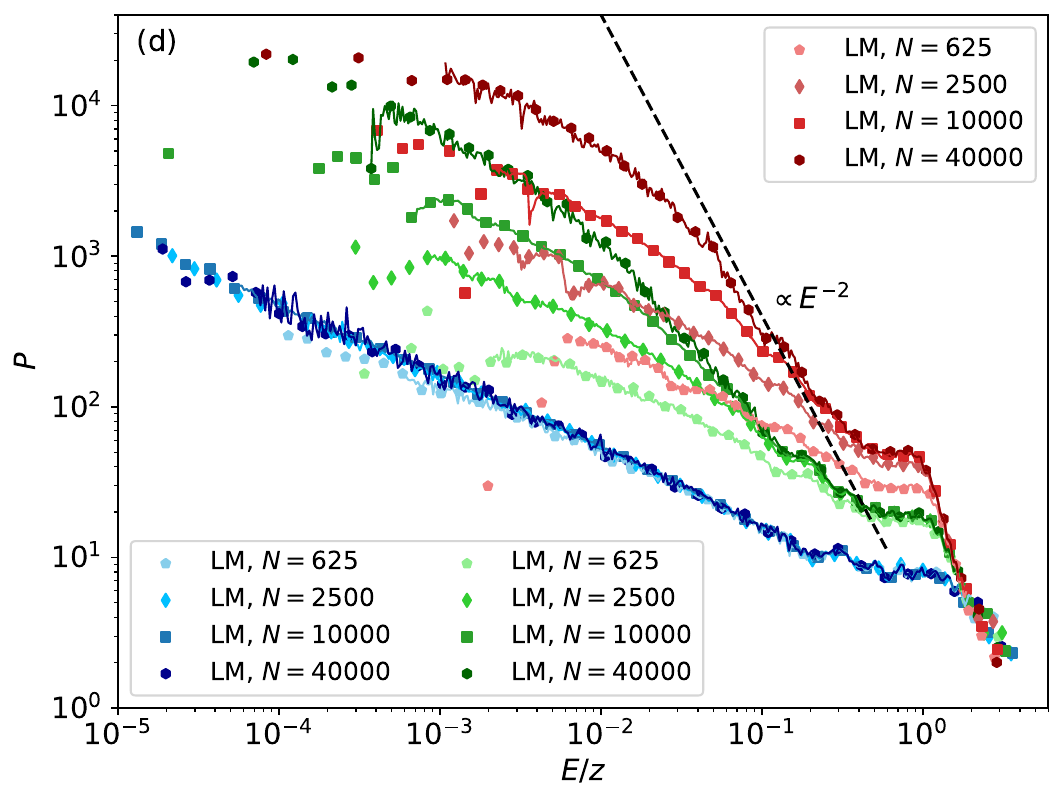}
    \end{subfigure}
    \caption{Dependence of the participation ratio on the energy for the continuous part of the spectrum. On the left $z$ is varied for constant $N=10^4$, and on the right the system size $N$ changes for fixed $z=6$ (green) and $z=8$ (red). In the top row, the AM is shown and in the bottom row the LM. The expected asymptotic power law $P\propto E^{-2}$ for the LM is shown as dashed line. The horizontal dashed lines indicate $P=2L/3$ and $P=4L/9$, i.e. the participation ratios of sinusoids in 1D and 2D, respectively. An ensemble of 1000 systems was simulated apart from $N=4\times 10 ^4$ where it was 10 realisations.}
    \label{fig:EvP}
\end{figure}

In the AM, the participation ratio increases with the energy until $E=1$, where there is a funnel-shaped minimum, and then decreases again (see Fig.~\ref{fig:EvP}(a)).
There is a minimum participation ratio which becomes larger for bigger values of $z$ and is approximately the same for positive and negative energies.
The energy values are not distributed symmetrically around $E=1$ but have a larger tail towards negative energies which also increases with $z$.
With increasing $z$, the maximum participation ratio also increases and for $z=32$ there is a plateau near the $P$ value where the PPDF shows the finite-size peak, indicating that this plateau is a finite-size effect and that the maximum would be at larger values of $P$ is the system size was larger, as can also be seen in Fig.~\ref{fig:EvP}(b). As we have argued above, all eigenmodes in the AM are fully localised for infinite system size, but the corresponding value of the maximum of $P$ is not yet visible in our data. 

In the LM, the participation ratio increases with decreasing energy and reaches a plateau between $P=\frac{2N}{3}$ and $P=\frac{4N}{9}$.
The former is the participation ratio of 1D sine functions, i.e.~of eigenfunctions of the ordered system that are periodic in one dimension and constant in the other, and the latter of eigenfunctions of a regular system that are periodic in both dimensions. As we have seen above (Fig.~\ref{fig:eigenvectors_LM}), the lowest-energy eigenmodes of the RGG show a strong periodic structure in one direction and a weak amplitude modulation in the other.  
For $z\lesssim5$, this plateau is not reached, because a considerable number of sites of the graph sit in small components.
On the right-hand end side of the plot, finite-size effects lead to a bump for large $z/N$. In the data for $z=16$ and 32, the steep drop followed by a minimum 
is again due to the steep decrease of the DOS in this energy region.
The slope of the function $P(E)$ increases with increasing system sizes (see Fig.~\ref{fig:EvP}(d)).
We expect the limiting slope of $-2$ based on the scaling behaviour of density of states and of the probability distribution of the participation ratio, see Eq.~\eqref{powerlaw} above and the surrounding text. 

\section{Discussion}

Although the Laplacian and adjacency matrix contain the same information, their density of states and the localisation properties of their eigenmodes are qualitatively different. The Laplacian matrix is related to diffusive dynamics on the graph, which means that there is a conservation law. The model must therefore have system-spanning eigenmodes that allow diffusive relaxation of a disequilibrium between the two halves of the system. All eigenvalues of the Laplacian matrix are non-negative, while the adjacency matrix has positive and negative eigenvalues. For a bipartite system, there is a one-to-one mapping between eigenmodes with positive and negative eigenvalues, but since RGGs are not bipartite, the spectrum extends farther to the left (i.e. negative eigenvalues) than to the right in our data. Large positive eigenvalues require opposite signs of the eigenfunction entries on neighbouring sites (because we evaluate the eigenvalues of the negative of the  adjacency matrix), a condition which cannot be perfectly satisfied if the graph is not bipartite. 

Another difference between the data for the Laplacian model (LM) and adjacency model (AM) is due to the fact that orbits (i.e.~eigenmodes that are fully localised on a few lattice sites) all have the eigenvalue 1 (or 0) for the adjacency matrix, but can have any nonnegative integer value for the Laplacian matrix. For smaller values of $z$, where orbits are more frequent, this leads to a deep funnel around 1 in the density of states of the AM, and to a series of slight funnels in the LM. 

The data change strongly as $z$ is changed. For $z=4$, the RGG has no system-spanning component, and the cutoff in the component size distribution leads to a corresponding cutoff in the participation ratio. 
For $z=6$, parts of the system-spanning component are only weakly connected to other sites, leading to a stronger heterogeneity of the graph than for larger $z$. 
For the LM, system-spanning modes must approach sine waves with large wavelengths for sufficiently large system sizes, when disorder averages well over the distance of a wavelength. 
For smaller $z$, good averaging requires larger wavelengths, which in turn requires larger system sizes. 
In this respect, the system-spanning modes of the LM show larger finite-size effects for smaller $z$. 
On the other hand, localised modes show larger finite-size effects in the LM and in the AM when $z$ is larger, because localisation can occur only over distances for which sites have sufficiently different neighbourhoods. 
We may call a region of the extension of this distance an effective lattice site. 
So the number of effective lattice sites is smaller for larger $z$. 
Furthermore, two-dimensional disordered systems have a phase transition to delocalisation at a disorder strength zero. 
This means that the localisation length becomes very large for larger $z$, making finite-size effects considerably more pronounced than in one-dimensional RGGs \cite{schaeferLocalizedDelocalizedModes2025}. 
A third effect of large $z$ is a pronounced peak in the density of states in the LM, leading to pronounced drops in the probability distribution of the participation ratio. 

Due to the conservation law, the LM shows eigenmodes of all sizes, resulting in a power-law tail of the probability distribution of the participation ratio when the system size is large enough that the wavelength of the eigenmodes span many effective sites. Our data show an approach to this power law with the exponent $-1.5$ for large system sizes, but much larger system sizes would be required to see this scaling over a larger range of values. On these large distances, the specific type of disorder associated with an RGG should not matter, so that the scaling behaviour of the large-wavelength modes of the LM agrees with that of diffusion on other two-dimensional systems with short-range correlated disorder, such as regular lattices with random coupling strengths. The above scaling considerations, which we employed to derive the exponent $-1.5$ are a straightforward generalization of the considerations we employed in the 1D case \cite{schaefer2025scaling}. In the same way, we can conclude that the number of system-spanning modes of the LM must scale as $\sqrt{N}$ (when $z$ is fixed): We have argued above that the spatial extension of large-wavelength eigenmodes of the LM is proportional to $\lambda^2$ in each of the two dimensions. For the sake of this argument, let us scale the linear system size $L$ such that $N \propto L^2$, i.e.~we keep the node density fixed when approaching the thermodynamic limit. Eigenmodes with a wavelength such that $\lambda^2\ge L$ show strong finite-size effects and span the entire system. Possible wavelengths must satisfy $\lambda = L/n$ with an integer $n$. The number of $n$ values associated with system-spanning modes is therefore proportional to $ \sqrt L$ in each direction, leading to a number of system-spanning modes proportional to $L$, which in turn is proportional to $\sqrt N$. This means that in 2D RGGs the proportion of system-spanning eigenmodes of the Laplacian matrix decreases with system size as $1/\sqrt N$ (when $z$ is fixed), i.e., the proportion of system-spanning eigenmodes vanishes in the thermodynamic limit, where all (or almost all) eigenmodes must be localised. 

To conclude, by combining a variety insights from the Anderson model family and 1D RGGs, we could explain qualitatively the spectra of the Laplace and adjacency matrices and the localisation properties of the eigenmodes. System size, component size distribution, network motifs and mean degree all affect the data, and they must be taken into account when trying to understand the spectra of real-world RGGs. 

\section{Data Availability}
The data for this paper will be publicly available at 10.5281/zenodo.19347464.

\section{Funding}
This work was supported by Deutsche Forschungsgemeinschaft [Dr300/15-1].

\printbibliography
\end{document}